\documentclass[showkeys,preprint]{revtex4}
\usepackage{graphicx}
\usepackage{amsmath}
\usepackage{color}
\usepackage{amssymb}
\usepackage{bibentry}
\usepackage{amsfonts}
\usepackage{hyperref} 
\hypersetup{
    colorlinks,%
    citecolor=blue,%
    filecolor=black,%
    linkcolor=black,%
    urlcolor=black}

\begin{document}
\title{Probing Atomic Structure and Majorana  Wavefunctions  in Mono-Atomic Fe-chains on Superconducting Pb-Surface}
\author{R\'emy Pawlak$^1$}
\author{Marcin Kisiel$^1$}
\author{Jelena Klinovaja$^1$}
\author{Tobias Meier$^1$}
\author{Shigeki Kawai$^1$}
\author{Thilo Glatzel$^1$}
\author{Daniel Loss$^1$}
\author{Ernst Meyer$^1$}
\affiliation{$^1$Department of Physics, University of Basel, Klingelbergstr. 82, 4056 Basel, Switzerland.}
\date{\today}

\maketitle
{\bf Motivated by the striking promise of quantum computation, Majorana bound states (MBSs)~\cite{Ettore1937} in solid-state systems~\cite{Alicea2012,Kitaev} have attracted wide attention in recent years~\cite{Sato_2010,oreg_majorana_wire_2010,Lutchyn2010,Yazdani2013,Pientka2013,Yazdani2014,Klinovaja2013,
RKKY_Simon,RKKY_Franz,Mourik2012,deng_observation_2012,das_evidence_2012,marcus_MF}. In particular, the wavefunction localization of MBSs is a key feature  and crucial for their future implementation as qubits~\cite{Alicea2012,Kitaev}. Here, we investigate the spatial and electronic characteristics of topological superconducting chains of iron atoms on the surface of Pb(110) by combining scanning tunneling microscopy (STM) and atomic force microscopy (AFM). We demonstrate that the Fe chains are mono-atomic, structured in a linear fashion, and exhibit zero-bias conductance peaks at their ends which we interprete as signature for a Majorana bound state.  Spatially resolved conductance maps of the atomic chains reveal that the MBSs are well localized at the chain ends ($\lesssim$ 25~nm), with two localization lengths as predicted by theory~\cite{Jelina2012}. Our observation lends strong support to use MBSs in Fe chains as qubits for quantum computing devices.}
\\
\indent
Majorana fermions are real solutions of the Dirac equation and by definition fermionic particles which are their own antiparticles~\cite{Ettore1937}. While intensely searched in particle physics as neutrinos,
Majorana fermions have recently been  predicted to occur as quasi-particle bound states
in engineered solid-state systems~\cite{Alicea2012}. Such systems not only offer the possibility to observe the exotic properties of such MBSs, but also open up an interesting playground for topological quantum computing~\cite{Kitaev,Alicea2012}. 
The fundamental ingredients to generate MBSs in semiconductor-superconductor heterostructures is to combine a spin texture with an $s$-wave superconductor (SC) allowing to create a superconducting state with effective $p$-wave  pairing, giving birth to a new state of matter--topological superconductivity~\cite{Alicea2012,Kitaev}. MBSs arise as zero-energy states lying in the superconducting gap and being spatially localized at the interfaces.\\
\indent
To generate spin textures, theoretical proposals have suggested to employ 
nanowires and magnetic chains with strong spin-orbit interaction~\cite{Sato_2010,oreg_majorana_wire_2010,Lutchyn2010,Yazdani2013,Pientka2013,Yazdani2014} or  with self-tuning RKKY interactions~\cite{Klinovaja2013,RKKY_Simon,RKKY_Franz}.
Until now, only few experimental works have reported successful observations of a zero-bias conductance peak (ZBP) via transport measurements in semiconductors interpreted as  MBS signature~\cite{Mourik2012,deng_observation_2012,das_evidence_2012,marcus_MF}, however, without addressing their spatial localizations in detail. Remarkably, Nadj-Perge {\it et al.}~\cite{Yazdani2014} recently studied the spatial and spectral resolutions of the MBS in Fe chains on superconducting Pb by STM and reported the striking observation
of a ZBP at the end of the atomic chains, as generically expected for MBSs~\cite{Kitaev}.
Experimentally, the induced proximity gap probed in the Fe chain is found to be very small ($\approx$ meV), while the exchange interaction  is in the eV range, implying on theoretical grounds a large localization length  of the MBS wavefunction, in contrast to the observation~\cite{Yazdani2014} have triggered interesting discussions on the physical origin of the ZBP~\cite{Dassarma2014,vonoppen2014}. It further implies that possible MBS in such chains may easily hybridize into conventional fermions, raising the question as to whether MBS in such Fe/Pb hybrid system will exhibit non-Abelian braiding statistics~\cite{Kitaev}. 
\\
\indent
\begin{figure*}[t!]
\centering
\includegraphics[height=6cm]{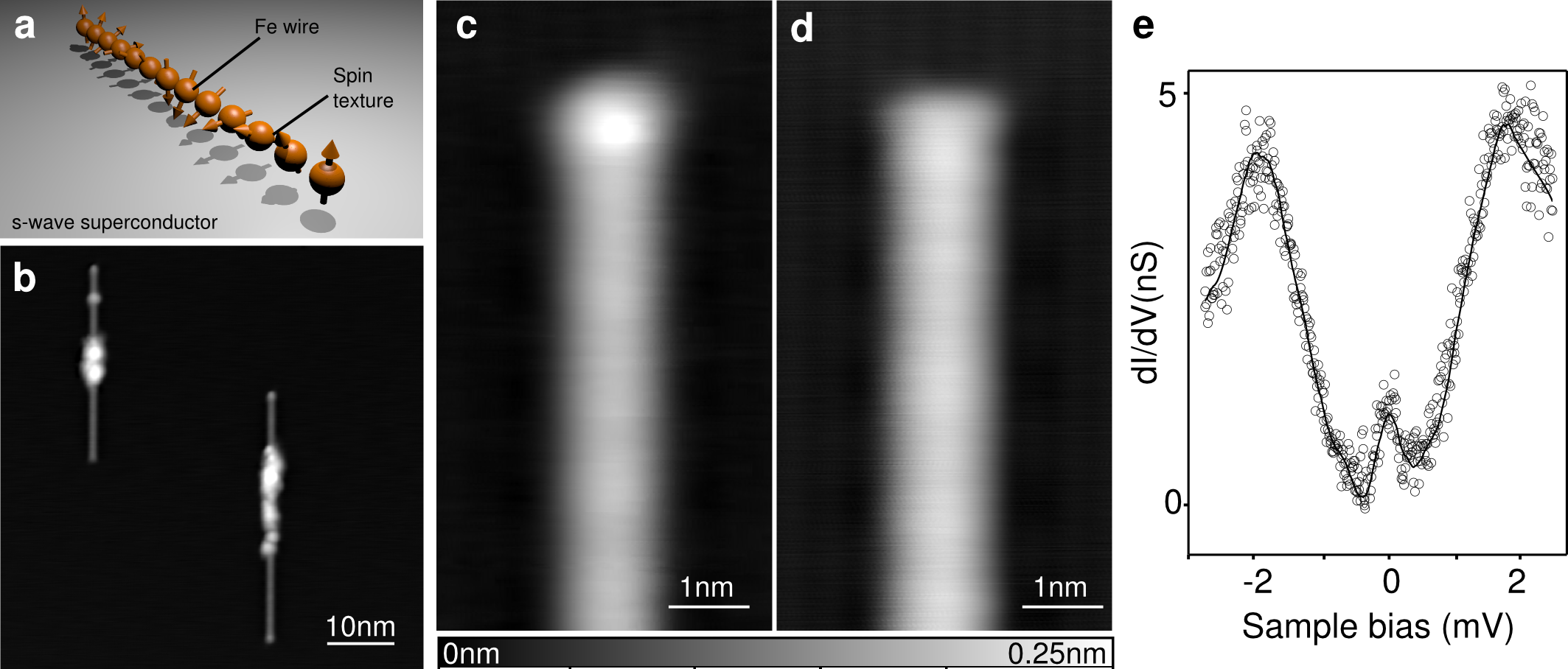}
\caption{{\bf Self-Assembled Fe-Chains on Pb(110)}. {\bf a,} Model of the experiment: topological superconducting phase arises when mono-atomic iron chains with spin texture are grown on an $s$-wave superconductor [here Pb(110)]. MBSs are then localized at the end of the chain and experimentally observed as ZBP in the conductance. {\bf b,} Topographic STM image of iron chains self-assembled on atomically-clean Pb(110). {\bf c-d,} Topographic STM images of two chain terminations with a difference of apparent height ($\approx$ 10~pm) at their ends, ($V_t$ = -10~mV, $I_t$ = 100~pA). {\bf e,} $dI/dV$ point-spectra obtained at the chain end in {\bf c} showing a ZBP. The proximity gap equal to $\Delta$ = 1.1~meV was measured with a metallic tip.
 } 
  \label{Fig1}
\end{figure*}
In this work, we go a decisive step further and combine STM and AFM measurements at low temperatures to  characterize the electronic properties and structure, respectively, of  superconducting Fe chains self-assembled on the surface of Pb(110), see Fig.~\ref{Fig1}{\bf a}. To engineer such systems, atomically cleaned surfaces of an $s$-wave superconductor, Pb(110), were prepared by a few sputtering/annealing cycles (see Supp. Info Fig.~S1) on top of which Fe atoms were deposited on the sample kept at $\approx$ 400~K. In such way, topographic STM images reveal the formation of long chains extending up to 70~nm (Fig.~\ref{Fig1}b). In agreement with ~\cite{Yazdani2014}, chains are partially covered by small clusters mainly localized around their centers (see Supp. Info Fig.~S2). Careful real-space observations reveal that numerous Pb vacancies are also created in the vicinity of the chains. We assume that the relative high temperature required for the self-assembly also promotes the diffusion of Pb atoms and their nucleation at the chain sides. During our study, several chains having lengths from 40 to 70~nm (Fig.~\ref{Fig1}b) were investigated and showed two distinct topographic signatures by means of STM at their ends (Fig.~\ref{Fig1} c and d).  They correspond to a variation of the apparent STM height of about 10~pm. Along the chains and independent of the termination types, no clear atomic periodicity has been observed by STM (see Fig.~S3) which could unambiguously be related to the structure of the chain. We then systematically acquired conductance point-spectra at the two chain ends to identify the presence of a ZBP which we interprete as signatures for MBS. Figure~\ref{Fig1}e shows a typical single-point conductance spectrum at the end of the chain with end-state (Fig.~\ref{Fig1}c) obtained with metallic tip. The superconducting gaps measured with such tip are thus equal to $\Delta$ = 1.1~meV being the superconducting gap of lead. Therefore, we particularly focus on those chains which exhibit end-states, as they show a clear signature of zero-energy end states in contrast to chains such as shown in Fig.~\ref{Fig1}d (See details in Supp. Infos Fig.~S4).  We note that the lowest temperature of our experiments is 4.7~K which limits our spectral resolution to about 1.4~meV.  Although the energy gap in our spectra is broadened compared to \cite{Yazdani2014} due to the higher measurement temperature, a superconducting proximity gap of about = 1.1~meV is observed in the Fe chain, as well as a clear ZBP which we interprete as a signature of a MBS that has been predicted to occur in such systems due to various mechanisms~\cite{Yazdani2013,Klinovaja2013,RKKY_Simon,RKKY_Franz,Pientka2013,Yazdani2014}.\\
\indent
\begin{figure*}[ht!]
\centering
\includegraphics[height=11cm]{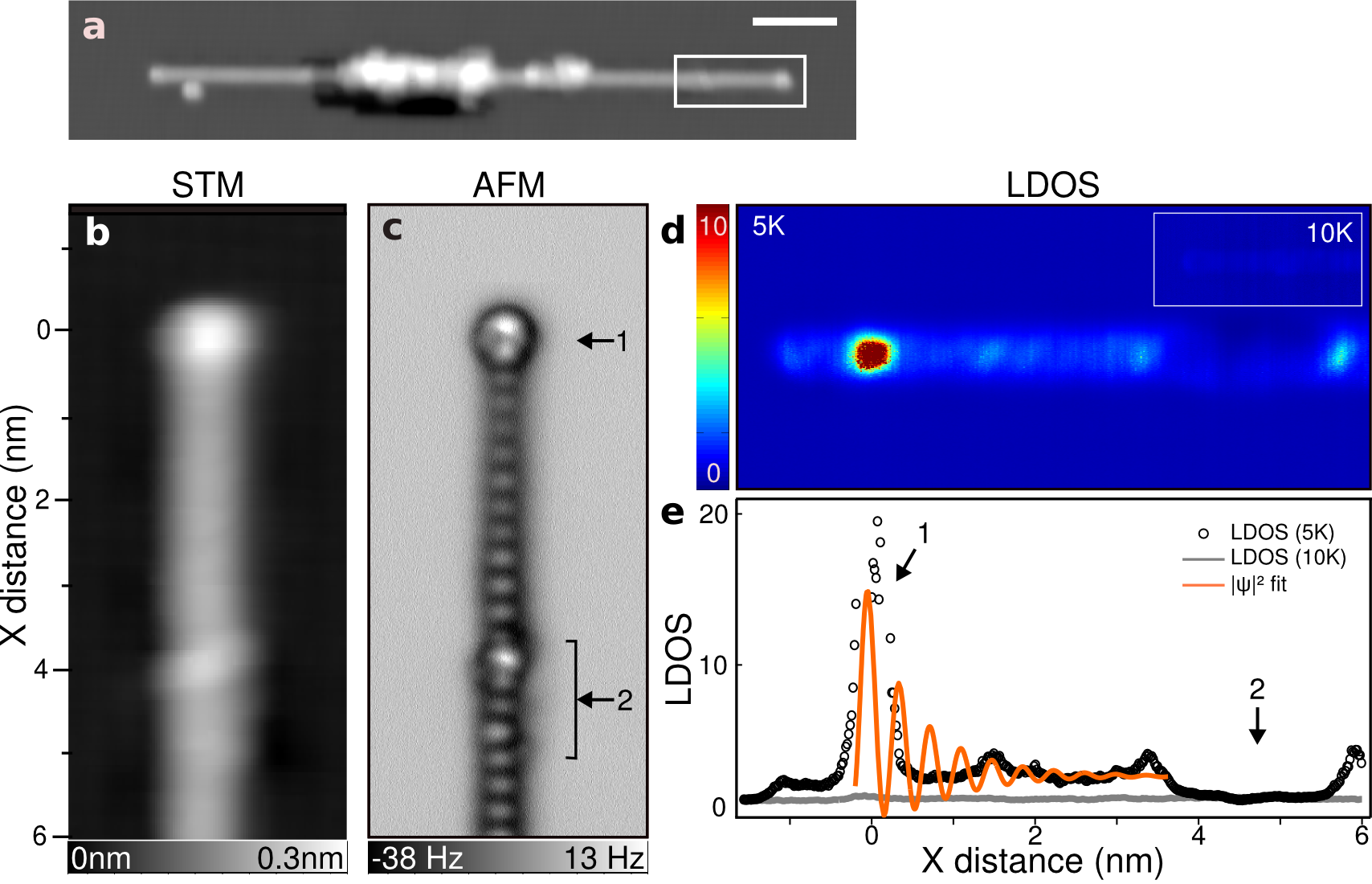}
\caption{{\bf Zero-Bias Electronic and Structural Characterization of the Fe Chain}. {\bf a,} STM topographic image of the Fe chain end ($V_t$ = -11~mV, $I_t$ = 80~pA, scalebar = 5~nm). The inset shows the section characterized in the rest of the figure. {\bf b,} STM topography at $V_t$ = 10~mV and $I$ = 10~pA. {\bf c,} Constant-height zero-bias AFM image of the chain end acquire at 5~K. In contrast to STM topography, the AFM image clearly resolves the atomic structure of the Fe chain. Each protrusion corresponds to single Fe atoms aligned in a straight fashion along the Pb(110) rows (Fe inter-atomic distance = 0.37~nm). In {\bf c} site {\it 1} shows the MBS and {\it 2} the presence of slightly misaligned atoms along the chain. {\bf d,} Zero-bias normalized conductance maps $\propto$ LDOS acquired at 5~K. At site $1$  where the MBS is localized, a zero-bias conductance peak is clearly observed at 5~K and vanishes at $10~K$ as a result of the suppression of the superconductivity. The LDOS obtained at 10 K shown in the inset is homogeneous along the chain whereas weak oscillations are observed at 5~K due to the MBS wavefunction decay. {\bf e,} Comparison of LDOS profiles extracted along the chain at 5~K (black dots) and 10~K (grey dots). The orange curve shows a  fit of the LDOS with the theoretical MBS wavefunction composed of two localization lengths.} 
  \label{Fig2}
\end{figure*}
{\it Electronic and Structural Characterization along the Mono-atomic Fe chain.} 
A ZBP localized at the end of the chains is one of the hallmarks of a MBS~\cite{Yazdani2014}. Alternatively, however,  such a ZBP can arise from magnetic impurities such as Shiba (near) mid-gap states in the vicinity of single adatoms~\cite{Yazdani1997,Ma2009}, molecules~\cite{Franke2011,Franke2013} or disorder effectss. To further identify the origin of the observed zero-energy subgap 
states and attribute it to a MBS, we compared STM and AFM imaging obtained at the atomic scale below and above the superconducting transition temperature 
(Fig.~\ref{Fig2}).\\
\indent
Figure~\ref{Fig2}b shows a close-up STM topography of the investigated part of a long chain. The chain exhibits the topographic signature of an electronic state at its ends as well as a weak perturbation of the electronic density along the chain. This last feature  might be caused by the presence of defects below or within the chain. Howewer, since STM reflects the electronic density between tip and sample, the `true' atomic structure can be masked by delocalized electronic states of the system~\cite{Gross2009,Pawlak2011}. As we noticed for the present system, the STM topographic data can lead to important misinterpretations of the atomic structure of the chain.
\\
\indent
To unambiguously address this issue, we employed the AFM imaging technique which is rather insensitive to the delocalization of electronic states close to the Fermi level.  Figure~\ref{Fig2}c shows a constant-height zero-bias AFM image obtained at 5~K reflecting the true atomic structure of the chain. Each protrusion corresponds to a single Fe atom which are all aligned and centered between the atomic rows of the Pb(110). The chains are mono-atomic and strictly ordered in a linear fashion. The Fe inter-atomic distance is equal to 0.37~nm (see profile in Supp. Infos. Fig.~S4), which is in good agreement with the atomic lattice along the Pb rows (about 0.35~nm). This suggests a high commensurability of the chain periodicity with the underlying substrate (the mismatch is about 0.6~$\%$) in contrast to the STM observation (see Supp. Infos. Fig.~S3).\\
\indent
The site marked by the arrow {\it 1} in Fig.~\ref{Fig2}c ($X$ = 2~nm) corresponds the MBS location and shows a less negative $\Delta f$ of about -2~Hz compared to the rest of the chain ($\Delta f$ background = -20~Hz, see profile Fig.~S3). This $\Delta f$ variation is localized at the last two atoms of the chain and corresponds to a less attractive force regime between tip and sample. The AFM map (Fig.~\ref{Fig2}c) shows a perfect round-shape halo of 0.8~nm in diameter as a spatial signature of this peculiarity. Within the halo, the last atoms are visible and perfectly aligned with the rest of the chain thus excluding atomic disorder or corrugation effects. The decrease of the attractive forces at such close tip-sample distances implies additional contributions of repulsive forces between tip and sample. Tuning fork AFM is known to be sensitive to such forces which usually originate from Pauli exclusion principle between the electronic wavefunctions of the tip and the surface~\cite{Gross2009}. Given that two end MBSs can host at most one fermion, it is plausible that a similar repulsive Pauli effect is responsible for the observed halo in the force maps. In comparison, site {\it 2} shows a less pronounced $\Delta f$ variation compared to site {\it 1} which coincides with slightly misaligned Fe atoms with respect to the chain structure.  The $\Delta f$ is less negative by $\approx$ -10~Hz compared to the chain. We further point out the absence of any halo at site $2$ and conclude that the $\Delta f$ variation is imputed to atomic disorders, most likely due to local topographic variations. We note  that  we also investigated chains without end-states and no such force contributions were observed (see Supp. Infos Fig. S5).\\
\indent
To further support our observation of  MBS, we compared normalized conductance maps, i.e. $(dI/dV)/(I/V)$ $\propto$ local density of state (LDOS) between tip and sample at the Fermi level, obtained at the same locations and at about 5~K and 10~K, respectively. Fig.~\ref{Fig2}d  shows the $LDOS(x,y)$ map at 5~K and reveals a clear ZBP, ascribed to the MBS, localized at the end of the chain as determined by the AFM data. A weak modulation of the LDOS is also observed along the chain which we attribute to the decay of the MBS wavefunction. At site {\it 2}, the LDOS is almost zero and might be due to weak magnetic perturbations induced by  lattice disorder~\cite{Yazdani1997,Franke2011,Franke2013}. In order to suppress the superconducting state of the system and thus to enforce the disappearance of the MBS, we  measured the same chain above the critical temperature of lead ($T$ $\geq$ $T_c = 7.2$~K). The corresponding normalized $LDOS(x,y)$ map (see inset Fig.~\ref{Fig2}d) shows a homogenous LDOS along the chain which drastically contrasts to the one at 5~K. By comparing the LDOS profiles taken along the chain at 5 and 10~K (black and grey dots in Fig.~\ref{Fig2}e), the ZBP has completely disappeared at 10~K (site {\it 1}) as well as the oscillations along the chain due to the suppression of topological superconductivity. Since no external magnetic field is applied~\cite{Yazdani2014}, we clearly address the interplay between superconductivity and the ZBP observed in our data which provides a strong evidence of the MBS presence in this system.\\
\indent
\begin{figure}[t]
\centering
\includegraphics[height=9.9cm]{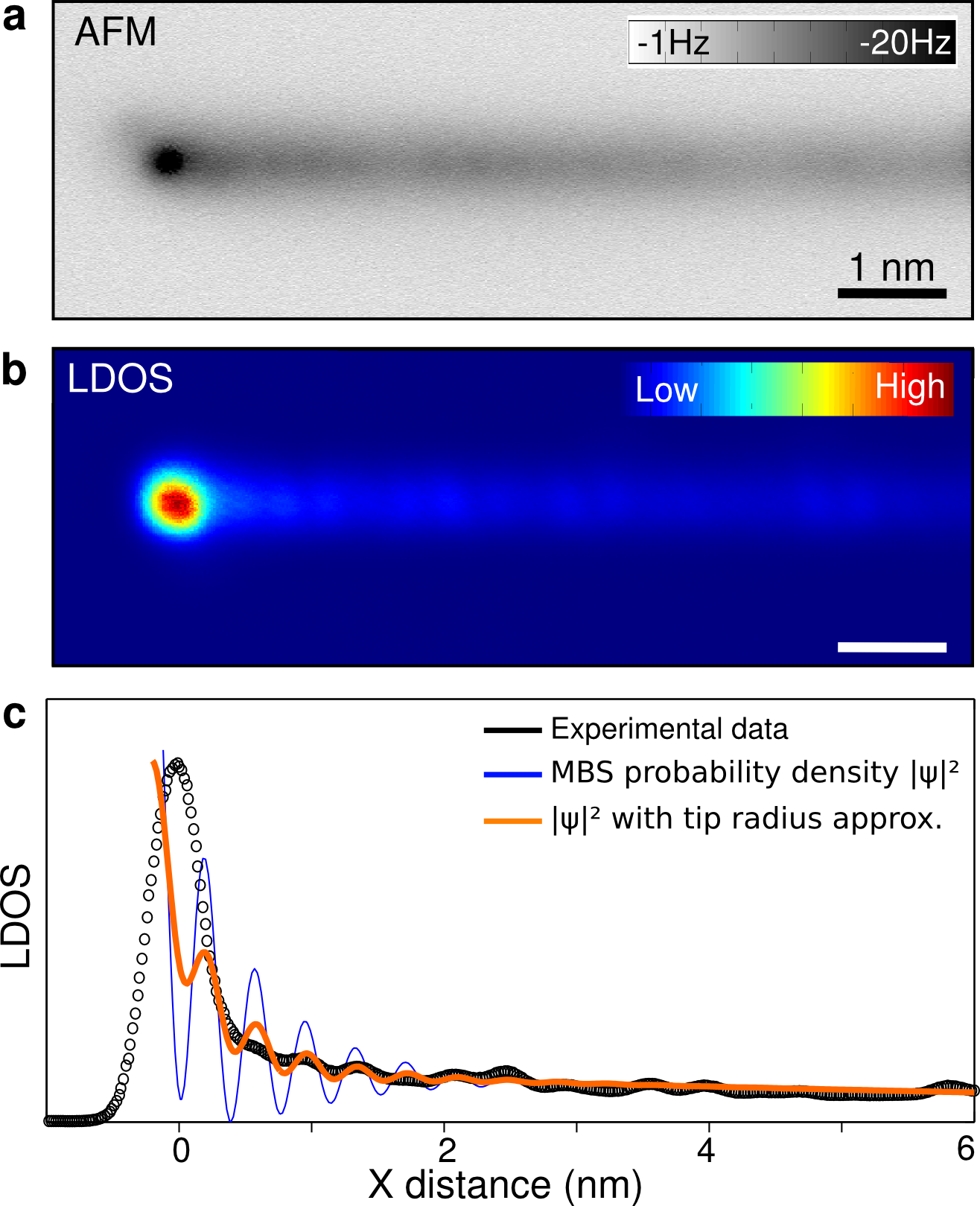}
\caption{{\bf Majorana Localization Lengths.}  {\bf a}, Constant-height zero-bias AFM image of the defect free chain. {\bf b}, Zero-bias normalized LDOS of the chain hosting a MBS. {\bf c}, $dI/dV(x)$ profile (red curve) taken along the chain revealing the localization lengths of the zero-bias conductance peak ($x$ = 0). The blue curve corresponds to the probability density $\vert \psi \vert^2$ of a Majorana Bound state with two localization lengths $\xi_1$ $\approx$ 22 nm and $\xi_2$ $\approx$ 0.72 nm. They correspond to 59 and 2 atomic sites respectively. The orange curve approximate the $\vert \psi \vert^2$ wave-function by considering the effect of the tip radius $d$ = 0.17 nm.} 
  \label{Fig4}
\end{figure}
{\it The} MBS {\it Localization Lengths.}
To further exclude  other explanations of the ZBP~\cite{Ternes2006}, we next investigate another strong fingerprint of a MBS, namely its wavefunction and associated localization lengths~\cite{Jelina2012}. For this we assume that the Fe chain with a proximity gap is driven into the topological phase  by a spin texture which gives rise to a helical field  (see Fig. 1{\bf a}). Although Nadj-Perge {\it et al.}~\cite{Yazdani2014} measured the chain magnetization and concluded to a ferromagnetic behavior, the precise value of the magnetization is however not known and could be substantially away from full ferromagnetic order. In turn, this does not exclude helical order since the helix can be around the magnetization axis with some small angle. Moreover, we wish to point out that, recently, such a helical state has been experimentally reported in a similar 1D-system~\cite{Menzel2014}.  Such helical fields can either follow  from spin orbit interaction  combined with Zeeman fields~\cite{Lutchyn2010,oreg_majorana_wire_2010,Pientka2013} or exchange fields~\cite{Yazdani2013,Yazdani2014} or from a `self-tuning' RKKY interaction between the Fe spins~\cite{Klinovaja2013,RKKY_Simon,RKKY_Franz}. These two mechanisms are related by a spin-dependent gauge transformation and thus mathematically equivalent~\cite{Braunecker_2010}.
We further assume that the pitch of the helix is much larger than the lattice constant of the Fe chain and that the helical field and the spin orbit interaction is sufficiently strong.
In this case, the MBS wavefunction is a superposition of contributions coming from different extrema in the spectrum, which results in two different localization lengths for a single MBS~\cite{Jelina2012}. 
Assuming a semi-infinite chain with one MBS at each end (without overlap), these localization lengths are determined by the corresponding gaps and given by
\begin{eqnarray}
\xi_1  = \frac{\hbar v_F}{\Delta}, ~~ \xi_2 = \frac{\hbar v_F}{\Delta_m - \Delta}.
\label{equ1}
\end{eqnarray}
The MBS probability density is then defined as
\begin{eqnarray}
\vert \psi(x)\vert^2&=& \frac{1}{N} \{e^{-2x/\xi_1} + e^{-2x/\xi_2}  \nonumber\\
 &-&2e^{-x(1/\xi_1+1/\xi_2)}\cos(2k_Fx)\},
\label{equ2}
\end{eqnarray}  
where $N$ is the normalization constant, $\xi_1$ and $\xi_2$ are the localization lengths, $\Delta$ and $\Delta_m$ are the proximity gap and the helical field gap of the chain (which depends on the exchange coupling constant), respectively, and $k_F$ is the Fermi wavevector of the chain.  The topological phase of the chain is reached when $\Delta_m$ $>$ $\Delta$~\cite{Klinovaja2013}.\\
\indent
The black curve shown in Fig.~\ref{Fig2}e show the $LDOS$ profile extracted along the Fe chain at 5~K. The orange curve represents a tentative fit with the theoretical wavefunction $\vert \psi \vert^2$ giving  $\xi_1\approx 110~nm$ and  $\xi_2 \approx 0.75~nm$. We remark that $\xi_2$, constituting the short localization of the MBS, is about the same as the diameter of the halo observed by AFM in Fig.~\ref{Fig2}c. However, the second localization length $\xi_1$ is only indicative as it exceeds the chain length and is presumably affected by the defect states present in this chain.\\
\indent   
To improve on such issues, we systematically investigated defect-free chains hosting MBS identified by both AFM and conductance measurements. Figures~\ref{Fig4}a and b show the zero-bias constant-height AFM maps and the normalized $LDOS(x,y)$ of such defect-free chains below $T_c$. Both data were acquired at slightly larger tip-sample separations ($\approx$ +50~pm) compared to Fig.~\ref{Fig2}c. For that reason, the AFM image do not resolve the atomic lattice along the chain. However, both the dark halo (Fig.~\ref{Fig4}a) and a ZBP in the conductance map Fig.~\ref{Fig4}b remains present at the chain end testifying the MBS. As previously discussed, a non-negligible LDOS signal along the chain is detected in the LDOS map Fig.~\ref{Fig4}b and decays with respect to the chain end. We extracted the period of the oscillations by performing a Fourier transformation (FT) of the conductance map (See Supp. Infos. Fig.~S7), which  gives us the Fermi wavevector $k_F$ equal to 8.3~nm$^{-1}$, {\it i.e.}  a Fermi wavelength of $0.76$~nm. The data was fitted with Eq.{\it (2)} by keeping constant the experimentally deduced parameters $k_F$. Fig.~\ref{Fig4}c shows the experimental $LDOS(x)$ profile extracted along the chain (black dots).  As shown by the superimposed blue curve, a remarkable agreement with the raw data periodicity is obtained allowing the extraction of the two localization lengths $\xi_1$ and $\xi_2$. These localization lengths are then found to be 22~nm and 0.75~nm, respectively, and correspond to about 59 and 2 atomic sites with respect to the chain lattice. Several data sets were analyzed this way always showing the same values for these localization  (see Fig.~S5). In addition, the proximity gap associated with $\xi_2$ is found to be $\Delta_m\approx$ 33~meV.\\
\indent
To account for the effect of the tip on the wavefunction measurement, we modify the formula for the probability  density $\vert \psi \vert^2$  by including a broadening effect resulting from a tip size effect (see additional text in Supp. Info.). This approximation considers a symmetric apex having metallic character and giving the modulus squared of the wavefunctions at the Fermy energy. The orange curve of Fig.~\ref{Fig4}b shows the result of such an approximation using a tip apex of 0.17~nm and demonstrates that such a broadening effect is sufficient to reasonably reconstruct the experimental data. We think that the use of lower measurement temperatures and $p$-wave STM tips~\cite{Leo2011} in future experiments might help to resolve more precisely the MBSs wavefunction.\\
\indent
In conclusion, our results validate the existence of MBSs at the end of atomic iron chains on superconducting lead. Atomic-scale AFM imaging shows that Fe adatoms form mono-atomic and straight chains on the Pb(110) surface, where zero-bias conductance peaks emerge at their extremities which we interpret as signature for MBS. These conductance peaks do not survive the suppression of superconductivity when increasing the sample temperature, in accordance with the expected behavior of MBSs. Comparison between current and force channels further demonstrates that the AFM imaging is sensitive to the MBS observed as an additional force contribution. Furthermore, we spatially characterized the localization of the MBS wavefunction which is composed of two localization lengths extending up to about 60 atomic sites. This suggests that the relatively short localization length of the MBS in such hybrid Fe/Pb systems arises from weak magnetic coupling of the chain atoms and might be tuned by weak external magnetic fields~\cite{Klinovaja2013,Jelina2012}.

\section*{Method}
{\bf Sample Preparation}
 Single Pb(110) crystal were provided by Mateck GmbH. After an {\it ex-situ} chemical treatment with hydrogen peroxide and acetic acid solution, the sample were atomically cleaned under ultrahigh vacuum by several cycles of sputtering and annealing. Iron atoms were deposited from a heated e-beam evaporator with a rate of $\approx$~0.07~ML/min on the surface and annealed at $\approx$~400~K in order to promote the formation of chains. \\

{\bf Scanning Probe Microscopy.}
The SPM measurements were realized with a low-temperature STM/AFM microscope (Omicron Nanotechnology GmbH) based on a tuning fork sensor ($f_0$ $\approx$ 25~kHz, $k$ $\approx$ 1800~N/m) and operated at $\approx$ 5~K in ultrahigh vacuum (UHV). All STM images were recorded in the constant current mode with the bias voltage applied to a bulk tungsten tip. The very-end of the tip apexes were prepared into a clean Cu(111) surface by gentle indentations. Conductance measurements were done at constant-height and zero-bias using the lock-in technique ($f$ = 570~Hz, $A$~=~200~$\mu$V) with the feedback loop opened at $I$ = 100~pA, V = 10~mV.  AFM images were performed in the constant-height mode at zero-bias with oscillation amplitudes of $A$ = 50~pm. The AFM images were obtained in the constant-height mode with a tuning fork sensor oscillating with amplitude of 50~pm. The variation of the resonance frequency, $\Delta f$, resulting from site-dependent interaction forces between tip and sample are dynamically tracked using a phase-lock loop while the probe travels over the surface.\\

\section*{Acknowledgements}
This work was supported by the Swiss National Science Foundation (SNF),  NCCR QSIT, Swiss Nanoscience Institute (SNI), Polish-Swiss Project PSPB-085/2010 and the EU Cost action MP1303. \\

\section*{Author contributions}
R. P., D. L. and E.M. conceived and designed the research. R. P., M. K. and T. M. prepared the samples. R. P. and M. K. performed the measurements. J. K. and D. L. provided theoretical support. R. P., M. K., E. M., D. L. and J. K. analyzed the data and wrote the manuscript with contributions from all authors.

\section*{Additional information}
Supplementary information is available in the online version of the paper.
Reprints and permissions information is available online at www.nature.com/reprints. Correspondence and requests for materials should be addressed to R. P., D. L. $\&$ E. M.

\section*{Competing financial interests}
The authors declare no competing financial interests.


\begin{thebibliography}{10}

\bibitem{Ettore1937}
Majorana,  E. Teoria simmetrica dell'elettrone e del positrone. {\it Nuovo Cimento} {\bf 14,} 171 (1937).

\bibitem{Alicea2012}
Alicea, J. New Directions in the Pursuit of Majorana Fermions in Solid State Systems. {\it Rep. Prog. Phys.} {\bf 75,} 07501 (2012).

\bibitem{Kitaev} 
Kitaev, A. Y. Unpaired Majorana fermions in quantum wires. {\it Phys.-Usp.} {\bf 44,} 131 (2001).

\bibitem{Sato_2010}
Sato, M., Takahashi, Y. $\&$ Fujimoto, S. Non-Abelian topological orders and Majorana fermions in spin-singlet superconductors. {\it Phys. Rev. B} {\bf 82,} 134521 (2010).

\bibitem{oreg_majorana_wire_2010}
Oreg,  Y.,  Refael,  G. $\&$ v. Oppen, F. Helical Liquids and Majorana Bound States in Quantum Wires. {\it Phys. Rev. Lett.} {\bf 105,} 177002 (2010).

\bibitem{Lutchyn2010}
Lutchyn, R. M., Sau, J. D. $\&$ Das Sarma, S. Majorana Fermions and a Topological Phase Transition in Semiconductor-Superconductor Heterostructures. {\it Phys. Rev. Lett.} {\bf 105,} 077001 (2010).

\bibitem{Yazdani2013}
Nadj-Perge, S., Drozdov,  I. K., Bernevig, B. A. $\&$ Yazdani, A. 
Proposal for realizing Majorana fermions in chains of magnetic atoms on a superconductor.
{\it  Phys. Rev. B}  {\bf 88,} 020407 (2013).

\bibitem{Pientka2013}
Pientka, F., Glazman, L.J. $\&$  v. Oppen, F. Topological superconducting phase in helical Shiba states. {\it Phys. Rev. B} {\bf 88,} 155420 (2013).

\bibitem{Yazdani2014} 
Nadj-Perge, S. {\it et al.} Observation of Majorana fermions in ferromagnetic atomic chains on a superconductor. 
 {\it Science} {\bf 346,} 602 (2014).

\bibitem{Klinovaja2013}
Klinovaja, J., Stano, P., Yazdani, A. $\&$ Loss, D. Topological Superconductivity and Majorana Fermions in RKKY Systems. {\it Phys. Rev. Lett.} {\bf 111,} 186805 (2013).

\bibitem{RKKY_Simon} 
Braunecker, B. $\&$ Simon, P. {\it Phys. Rev. Lett.} {\bf 111,} 147202 (2013).

\bibitem{RKKY_Franz}
Vazifeh, M. $\&$ Franz, M. {\it Phys. Rev. Lett.} {\bf 111,} 206802 (2013).



\bibitem{Mourik2012}
Mourik, V. {\it et al.}
Signatures of Majorana fermions in hybrid superconductor-semiconductor nanowire devices. {\it Science} {\bf 336,} 1003 (2012).

\bibitem{deng_observation_2012} 
Deng, M.T. {\it et al.}
Anomalous zero-bias conductance peak in a Nb-InSb nanowire-Nb hybrid device. {\it Nano Lett.} {\bf 12,} 6414 (2012).

\bibitem{das_evidence_2012}
Das, A. {\it et al.} 
Zero-bias peaks and splitting in an Al–InAs nanowire topological superconductor as a signature of Majorana fermions. {\it Nat. Phys.} {\bf 8,} 887 (2012). 



\bibitem{marcus_MF} 
Churchill, H. O. H. {\it et al.}
Superconductor-nanowire devices from tunneling to the multichannel regime: Zero-bias oscillations and magnetoconductance crossover. {\it Phys. Rev. B} {\bf 87,} 241401(R) (2013).

\bibitem{Jelina2012}
Klinovaja J. $\&$ Loss, D. Composite Majorana fermion wavefuntions in nanowires, {\it Phys. Rev. B} {\bf 86,} 085408 (2012).

%
%




\bibitem{Dassarma2014} 
Dumitrescu, E., Roberts, B., Tewari, S., Sau, J. D. $\&$  Das Sarma, S. Majorana fermions in chiral topological ferromagnetic nanowires. {\it Phys. Rev. B} {\bf 91}, 094505 (2015).

\bibitem{vonoppen2014}
Peng, Y., Pientka, F., Glazman, L.I. $\&$ v. Oppen, F. Strong localization of Majorana end states in chains of magnetic adatoms. {\it Phys. Rev. Lett.} {\bf 114}, 106801 (2015).

\bibitem{Yazdani1997}
Yazdani, A., Jones, B. A., Lutz, C. P., Crommie, M. F. $\&$  Eigler, D. M.
Probing the local effects of magnetic impurities on superconductivity. {\it Science} {\bf 275,}  1767--1770  (1997).

\bibitem{Ma2009}
Ji, S. -H. {\it et al.}
Application of magnetic atom induced bound states in superconducting
gap for chemical identification of single magnetic atoms. {\it App. Phys. Lett.} {\bf 96,} 073113 (2010).

\bibitem{Franke2011}
Franke, K. J., Schulze, G. $\&$ Pascual, J. I. Competition of superconducting phenomena and Kondo screening at the nanoscale. {\it Science} {\bf 332,} 940--944 (2011).

\bibitem{Franke2013}
Heinrich, B. W.,	Braun, L., Pascual, J. I. $\&$ Franke, K. J. Protection of excited spin states by a superconducting energy gap. {\it Nature Phys.} {\bf 9,} 765--768 (2013).


\bibitem{Gross2009}
Gross, L., Mohn, F., Moll, N., Liljeroth, P. $\&$ Meyer, G. The chemical
structure of a molecule resolved by atomic force microscopy. {\it Science} {\bf 325}, 1110--1114 (2009).

\bibitem{Pawlak2011}
Pawlak, R., Kawai, S., Fremy, S., Glatzel, T. $\&$ Meyer, E. Atomic-scale mechanical properties of orientated C$_{60}$ revealed by noncontact atomic force microscopy. {\it ACS Nano} {\bf 54}, 6349 (2011).

\bibitem{Ternes2006}
Ternes, M {\it et al.} Subgap structure in asymmetric superconducting tunnel junctions. {\it Phys. Rev. B} {\bf 74,} 132501 (2006).


\bibitem{Menzel2014}
Menzel, M., Kubetzka, A., von Bergmann, Kristen $\&$ Wiesendanger, R. Parity effect in 120$^{\circ}$ spin spirals. {\it Phys. Rev. Lett.} {\bf 112,} 047204 (2014). 


\bibitem{Braunecker_2010}
Braunecker,  B., Japaridze, G. I., Klinovaja, J. $\&$ Loss, D. Spin-selective Peierls transition in interacting one-dimensional conductors with spin-orbit interaction. {\it Phys. Rev. B} {\bf 82,} 045127 (2010).

\bibitem{Leo2011}
Gross, L. {\it et al.}
High-resolution molecular orbital imaging using a {\it p}-wave STM Tip. {\it Phys. Rev. Lett.} {\bf 107,} 086101 (2011).
\end{thebibliography}
\end{document}